\documentstyle[epsf,rotate]{ioplppt}

\begin{document}
\jl{1}
\title{Twofold-broken rational tori and the uniform semiclassical
approximation}
\author{Henning Schomerus}
\address{Instituut-Lorentz, Universiteit Leiden,
P.\,O. Box 9506, 2300 RA Leiden, The Netherlands}

\begin{abstract}
We investigate broken rational tori consisting of a chain of four
(rather than two)
periodic orbits. The normal form that describes this configuration
is identified and used to construct
a uniform semiclassical approximation, which can be utilized to
improve trace formulae. 
An accuracy gain can be achieved even for the situation when two of the
four orbits are ghosts.
This is illustrated 
for a model system, the kicked top.
\end{abstract}

\pacs{ 05.45.+b, 03.20.+i, 03.65.Sq}
\maketitle

\section{Introduction}

Periodic orbits provide the skeleton of the dynamics of classical
Hamiltonian systems.
Generic dynamical systems display a mixed phase space,
consisting of islands of stability residing in chaotic seas.
The periodic orbits are neither grouped in families, like in 
integrable systems, nor are they well isolated and all unstable,
as for chaotic systems. One rather finds also stable periodic orbits
surrounded by islands of regular behaviour.
These islands of stability look locally like 
an almost integrable system, altogether with
a KAM structure of invariant tori and chains
of periodic orbits, remnants of rational tori
of a supposedly contiguous integrable situation
\cite{Lichtenberg:Lieberman:1992}.

In a semiclassical treatment
of the corresponding quantum system, clusters of
proximate orbits display a collective behaviour.
The bifurcations at the centre
of the island and their semiclassical treatment
have been addressed in a number of recent works, 
both for the generic variants
\cite{Ozorio:Hannay:1987,Ozorio:1988,Sieber:1996a,%
Schomerus:Sieber:1997,Sieber:Schomerus:1998}
as well as for classically
non-generic, but semiclassically still relevant
cases \cite{Schomerus:1997a,Schomerus:1998}.

In \cite{Tomsovic:Grinberg:1995} the class of near-integrable
systems has been addressed, and a uniform semiclassical
approximation for the most frequently encountered
broken rational tori
(consisting of a stable and an unstable periodic orbit)
was presented.  We will call these tori the `simple' tori.
Interestingly, the semiclassical approximation works reasonably
well even beyond the point where
the stable orbit becomes unstable \cite{Main:Wunner:1999}.
The same configuration of a stable and an unstable orbit is also typical
close (in parameter space) to most types of period-$n$-tupling
bifurcations at the centre of a stability island. For bifurcation number
$n\ge 5$,
for the island-chain scenario with $n=4$,
and also (as a consequence of a more complicated bifurcation scenario)
for $n=3$,
two satellite orbits are expelled from the centre.
At a certain distance from the bifurcation
the satellite orbits can be treated as isolated from
the central orbit;
however,
a collective semiclassical treatment of the two satellites is often still 
necessary, and this can be achieved by
using the abovementioned approximation for the simple torus.

Although encountered less frequently, there are situations where
a broken torus consists not of two, but a higher number
of periodic orbits. In the islands of stability
tori of this type appear especially 
at larger distance from a bifurcation.
In this work we study the twofold-broken rational torus,
consisting of two stable and two unstable periodic orbits.
It is described by a normal form which is obtained from the normal form
of the simple torus 
by including the second harmonic in an angular coordinate.
From the normal form we construct a uniform approximation that can be used to
improve semiclassical trace formulae.  Indeed, the relevance of this
configuration is much enhanced in the semiclassical context:
Here one has to consider also `ghost' orbits with complex coordinates
\cite{Kus:Haake:1993b}, and
even a simple broken torus can be affected by nearby ghosts, making a treatment
as a `pre-formed' twofold-broken torus advisable.
This is illustrated in a model system, a periodically
driven angular momentum vector (the kicked top), where we find a
configuration of four period-three orbits
which can be regarded as a twofold-broken rational torus.
A reduction of the error of the trace formula by a factor of about $2-3$
is found even when two of these satellites are ghosts.
An even higher accuracy gain is attained for so-called
`inverse-$\hbar$ spectroscopy'.

\section{Normal forms and uniform approximations}

We restrict the analysis to two-dimensional area-preserving maps.
(The results are also applicable to autonomous Hamiltonian systems with
two degrees of freedom.)
The quantum version of the map is generated by
the unitary Floquet operator $F$ which acts on
the vectors of a Hilbert space, mapping the space onto itself.
The semiclassical trace formula relates the traces $\mbox{tr}\,F^n$
to the classical periodic orbits.
Isolated orbits of primitive period $n_0$ give an additive contribution
\cite{Gutzwiller:1971,Gutzwiller:1990,Tabor:1983,Junker:Leschke:1992}
\begin{equation}
\label{eq:statphaseint}
C=A
\exp\left[\i\frac S\hbar - \i\frac \pi2\mu\right]
\end{equation}
with amplitude
\begin{equation}
A=
n_0
|2-\tr
M|^{-1/2}
\label{eq:amplitudes}
\end{equation}
to all traces $\mbox{tr}\,F^n$ with $n=n_0 r$ and integer repetition number $r$.
Besides the primitive period,
three classical quantities of the ($r$-th return of the) periodic orbit
enter, the action $S$,
the trace of the linearized $n$-step map
$M$, and the Maslov index $\mu$.

The expression (\ref{eq:statphaseint}) is derived by a stationary-phase
approximation and becomes inaccurate when orbits lie close together.
Then a collective treatment of the region $\Omega$
inhabited by the proximate orbits becomes
necessary. This can be achieved by introducing normal forms for a phase
function $\Phi$
and an amplitude function $\Psi$ into the more general expression
\cite{Ozorio:Hannay:1987,Ozorio:1988,Sieber:1996a}
\begin{equation}
\label{eq:pqint}
C_{\Omega}=\frac 1 {2\pi\hbar}
\int_{\Omega}\d\varphi'\;\d I\;
\Psi(\varphi',I)\exp\left[\frac \i \hbar
\Phi(\varphi',I)-\i \frac \pi 2 \nu\right].
\end{equation}
Here $I$, $\varphi$ are canonical polar (or cylinder) coordinates,
and $\nu$ is the Morse index.

The famous Poincar{\'e}--Birkhoff theorem states
that a perturbation of an integrable system causes tori with rational
winding number
to break into chains of alternating stable and unstable periodic points.
However, it does not give a quantitative criterion for the number of
distinct orbits that lie on this chain.
The simple broken torus is described by the normal form
\begin{equation}
S(I,\varphi')=S_0+I\varphi'-a I^2-b  \cos\varphi'
\end{equation}
for the generating function $S$, with constants $S_0$, $a$, and $b$. The
corresponding map $(I,\varphi)\to(I',\varphi')$,
implicitly given by
\begin{equation}
\varphi=\frac{\partial S}{\partial I},\quad
I'=\frac{\partial S}{\partial \varphi'},
\end{equation}
is the well-known standard map.
This normal form has been used in \cite{Tomsovic:Grinberg:1995}
to obtain a uniform semiclassical
approximation for the simple broken torus,
smoothly interpolating
between the two non-commuting classical and integrable limits 
($\hbar\to 0$ and $b\to 0$, respectively).

A more complete picture can be obtained when one
includes the second
harmonic in the angular variable $\varphi'$ and works with the extended
normal form
\begin{equation}
S(I,\varphi')=S_0+I\varphi'-a I^2 -b\cos\varphi'-c\cos(2\varphi'+2\varphi_0).
\label{eq:nform2}
\end{equation}
The strategy of including higher-order terms in normal forms has been
pursued before, with two different incentives. Firstly, the inclusion
of higher orders is a tool to equip a normal form with a sufficient number
of independent parameters. This can be necessary to account for all 
classical properties (stabilities and actions)
of the periodic orbits described by the normal form.
Secondly, the higher orders describe additional periodic orbits,
and hence more complicated configurations than the usual normal forms
\cite{Schomerus:1997a,Schomerus:1998,Sadovskii:Shaw:1995,%
Sadovskii:Delos:1996,Main:Wunner:1997}.
A semiclassical description often succeeds only when all orbits 
of an extended normal form are treated collectively.
Presently we aim at the inclusion of additional periodic orbits.

The periodic orbits satisfy the fixed point conditions
\begin{eqnarray}
&&I=\frac{\partial S}{\partial\varphi'}=
I+b\sin\varphi'
+2c\sin(2\varphi'+2\varphi_0),
\\
&&\varphi'= \frac{\partial S}{\partial I}=\varphi'-2a I,
\end{eqnarray}
resulting in $I=0$ and
\begin{eqnarray}
&&
b\sin\varphi' +2c\sin(2\varphi'+2\varphi_0)=0.
\end{eqnarray}
This condition amounts to finding the roots
of a fourth-order polynomial in $\sin\varphi'$.
Depending on the parameters $b$, $c$, and $\varphi_0$ there are
either four real solutions or two real and two complex solutions.
For $|b|<2|c|$ there are always four real solutions.
For $|b|>|c|$ only two real solutions are found.
When $|b/c|$ is fixed in the range $(1,2)$ and $\varphi_0$ is varied
one finds tangent bifurcations with two real solutions on one side
of the bifurcation and four real solutions on the other side.
The real solutions correspond to conventional periodic orbits
while the complex solutions are `ghosts'.
They are of no consequence for the classical dynamics but
can be important for the semiclassical description of the quantum system,
as has been shown in \cite{Kus:Haake:1993b} and as we shall
see once more below.

The joint contribution of the
orbits on the twofold-broken torus 
is found by introducing into Eq.\ (\ref{eq:pqint})
the normal form
\begin{equation}
\Phi=S_0-a I^2-b\cos\varphi'-c\cos(2\varphi'+2\varphi_0)
\label{eq:phi}
\end{equation}
[cf.\ Eq.\ (\ref{eq:nform2})] for the phase function and 
\begin{equation}
\Psi=1+d\cos(\varphi'+\varphi_1)+e\cos(2\varphi'+2\varphi_0)
\label{eq:psi}
\end{equation}
for the amplitude function.
The
stationary-phase limit of Eq.\ (\ref{eq:pqint}) is a sum
of four additive contributions of form
(\ref{eq:statphaseint}), each representing one orbit.
The four parameters $S_0$, $b$, $c$, $\varphi_0$ are determined
by matching the phases of each contribution to
the actions $S$ of the periodic orbits. The parameters $a$, $d$,
$e$, and $\varphi_1$ are fixed by the stability amplitudes
$A$. This strategy works also when two of the orbits are ghosts:
Phases and amplitudes become then complex,
but are related by complex conjugation,
and the number of real independent parameters remains unchanged.

Eqs.\ (\ref{eq:pqint},\ref{eq:phi},\ref{eq:psi})
represent a uniform approximation of the
joint contribution of the twofold-broken torus.
The integration over $I$ is readily carried out, which leaves us with a
one-dimensional strongly oscillating integral over the 
coordinate $\varphi'$. Numerically it is most conveniently evaluated by
the method of steepest descent, for which  the integration contour
is deformed into the complex plain. 
On the new contour
the integrand decreases
exponentially.
The new contour can also visit stationary points with
complex coordinates, i.\,e., ghost orbits.

\section{Numerical results}

We shall illustrate our findings,
and especially the relevance of ghosts,
for a situation encountered in a
model system, the dynamics
of a periodically driven angular momentum vector ${\bf J}$
(the kicked top \cite{Bibel}).
The components of ${\bf J}$ obey the usual commutation rules 
$[J_x,J_y]=iJ_z$ (and cyclic permutations).
The total angular momentum ${\bf J}^2=j(j+1)$ is conserved,
restricting the dynamics to the irreducible representations of the
angular-momentum algebra. The Hilbert space dimension is $2j+1$.
The effective Planck's constant is $1/(j+1/2)$ and the classical limit
is attained for $j\to \infty$.
We work here with a Floquet operator of the explicit form
\begin{eqnarray}
F&=&\exp\left[-\i k_z\frac {J_z^2}{2j+1}-\i p_z J_z\right]
\exp\left[-\i p_y J_y\right] \nonumber
\\
&&\times
\exp\left[-\i k_x\frac {J_x^2}{2j+1}-\i p_x J_x\right]
.
\end{eqnarray}
The dynamics consists of a sequence of linear
rotations by angles $p_i$ alternating with torsions
of strength $k_i$.  We hold the $p_i$ fixed
($p_x=0.3$, $p_y=1.0$, $p_z=0.8$) while varying
the control parameter
$k\equiv k_z=10k_x$. 
Complete semiclassical spectra of this system throughout the full
transition from integrable ($k=0$) to well-developed chaotic behaviour
($k\approx 10$) have been
presented in \cite{Schomerus:Haake:1997}. The system has also been used
to illustrate
uniform semiclassical approximations for various kinds of bifurcations 
 \cite{Schomerus:Sieber:1997,Sieber:Schomerus:1998,Schomerus:1997a}.

\begin{figure}
\epsfxsize6cm
\centerline{\rotate[r]{\epsffile{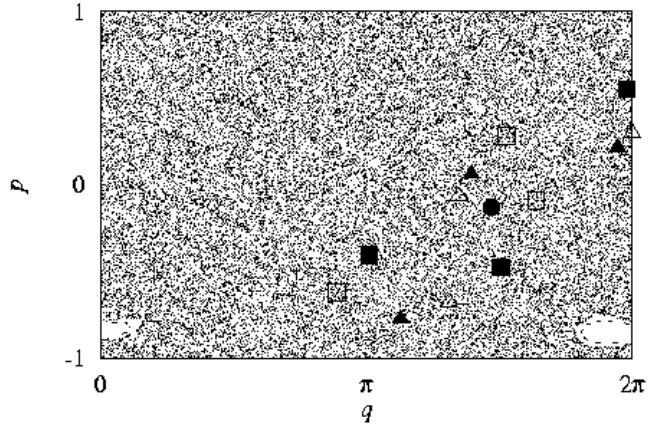}}}
\caption{
Phase space portrait of the kicked top with $k=4.0$.
The phase space is the unit sphere, parameterized here with the
azimuthal angle $q$ and the $z$-component $p$ (in conventional spherical
coordinates, $q=\phi$ and $p=\cos\theta$).
The circle indicates the central period-one orbit, the
other symbols indicate the positions of the four
period-three satellites (all orbits are unstable).
}
\label{fig:phasespace}
\end{figure}

\begin{figure}
\epsfxsize8cm
\centerline{\epsffile{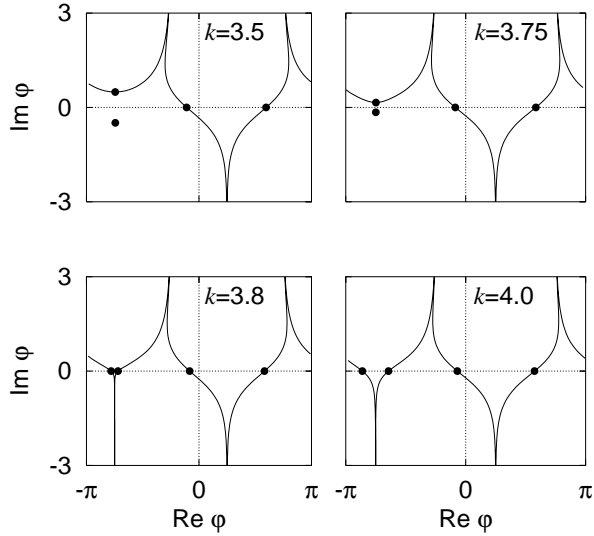}}
\caption{
Integration contours in the complex
$\varphi'$-plane on both sides of the 
tangent bifurcation in the kicked top at $k=3.7856$.
For each $k$ the normal-form coefficients
of the phase function $\Phi$
have been determined
from the actions of the periodic orbits.
The contours pass through the saddle points of the
normal form (circles) and proceed in the direction
of steepest descent (constant phase of $\Phi$).
}
\label{fig:contours}
\end{figure}

We concentrate on a particular
configuration of period-three orbits which suggests a treatment
as a twofold-broken torus. The configuration comes about in a sequence of three
bifurcations:  At $k=1.9715\ldots$ a pair of period-three satellites
is born in close vicinity to
a period-one orbit in the centre of a stability island.
At $k=1.9753$ a period-tripling bifurcation takes place where the stable
satellite collides with the central orbit.
On the other side of the
bifurcation the satellites form a  simple broken torus around the centre.
This sequence of
bifurcations can be described by extended normal forms
\cite{Schomerus:1997a,Schomerus:1998}, and the close neighbourhood of the two
bifurcations is not exceptional (see e.\,g.\ \cite{Mao:Delos:1992}).
As the control parameter $k$ is increased
further the satellites move away from the centre. At $k=3.7856$ another pair
of period-three satellites appears in a tangent bifurcation.
For $k<5$ the
four satellites form a configuration that resembles a twofold-broken torus.
Figure \ref{fig:phasespace} displays a phase space portrait for $k=4$, and
Figure \ref{fig:contours} shows steepest-descent contours in the
complex $\varphi'$-plane on both sides of the final bifurcation.

\begin{figure}
\epsfxsize8cm
\centerline{\epsffile{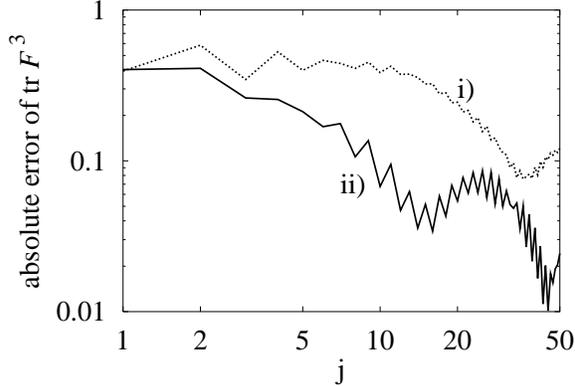}}
\caption{
Semiclassical error of ${\mbox tr}\, F^3$ for the kicked top
($k=3.0$) as
a function of $j$ in the two
approximations i), ii) which are explained in
the text.
}
\label{fig:trace}
\end{figure}

\begin{figure}
\epsfxsize8cm
\centerline{\epsffile{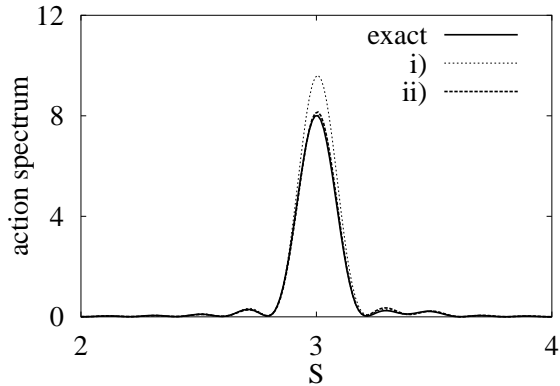}}
\caption{
Collective peak of the four satellites and the central orbit in the
action spectrum $|T(S)|^2$ for $k=3.0$. The semiclassical approximations
i) and ii) (see text) are compared to
the result of an exact quantum-mechanical computation.
}
\label{fig:actionspectrum}
\end{figure}

We evaluated $\mbox{tr}\,F^3$  at $k=3.0$ with the quantum number $j$
ranging from  $1$ to $50$.
At the given value of the control parameter only one of the  pairs of
satellites mentioned above has real
coordinates, and its 
distance to the centre of the stability island is already quite large.
The other two satellites are  still ghosts,
but their bifurcation is not far away.
This leaves us with the choice between two semiclassical approximations: 
i) We can group the two real satellites together with the central orbit
and treat the ghosts separately, or
ii) we can group the four satellites as a twofold-broken torus
and treat the central orbit separately.
(In principle, we could complicate matters even more and
group all the orbits together, but this is rather impractical and
beyond the scope of the present work.)
The semiclassical evaluation of the trace involves also five other orbits.
The error of the two approximations is shown in
Figure \ref{fig:trace}. The error of approximation ii) is about a
factor $2-3$ smaller than that of approximation i). Although the
accuracy gain is not dramatic, this result favours clearly
a treatment of the satellites as a pre-formed twofold-broken torus.

A somewhat more demanding test is aided by `inverse-$\hbar$ spectroscopy'
\cite{Tomsovic:Grinberg:1995,Wintgen}.
We consider
the discrete truncated Fourier analysis
\begin{equation}
T(S)=\frac{1}{32}\sum_{j=1}^{32}\mbox{tr}\,F^3(j)\exp[-\i (j+\case 12)S]
\end{equation}
of the trace with respect to the quantum number $j$.
The `action spectrum' $|T(S)|^2$ displays peaks at the  actions of the
periodic orbits that contribute to $\mbox{tr}\, F^3$.
Since accidental action degeneracies do not occur in the present example,
the quality of the semiclassical approximations i) and ii) can now be assessed
directly, without interference of the remaining orbits.
Figure \ref{fig:actionspectrum} shows the collective peak of the four satellites
and the
central orbit. Approximation ii) agrees almost perfectly with the exact result, 
while approximation i) overestimates the peak-height distinctively.

Of course, the uniform approximation presented here is not restricted to
the situation where two of the orbits are ghosts,
but is valid for four real orbits on the broken torus as
well. Indeed we find an improvement in the semiclassical 
accuracy of $\mbox{tr}\,F^3$ over the full range $2.8<k<5$.

\section{Conclusion}

In this paper a uniform approximation for a broken rational torus
consisting of four periodic orbits has been presented.
The approximation was tested in a model system where the phase-space is mixed,
and the tori are grouped around (but not too close to) a central orbit.
It can be expected that the approximation will be even more
useful in studies of globally near-integrable systems.

\ack

This work was supported by the DFG (Sonderforschungsbereich 237) and the
European Community
(Program for the Training and Mobility of Researchers).

\end{document}